# Inhibition of light emission in a 2.5D photonic structure

Romain Peretti, Christian Seassal, Pierre Viktorovich and Xavier Letartre

*Institut des Nanotechnologies de Lyon (INL), Université de Lyon, UMR 5270, CNRS-INSA-ECL-UCBL*

*Ecole Centrale de Lyon, 36 Avenue Guy de Collongue, 69134 Ecully Cedex, France*

Abstract

We analyse inhibition of emission in a 2.5D photonic structures made up a photonic crystal (PhC) and Bragg mirrors using FDTD simulations. A comparison is made between an isolated PhC membrane and the same PhC suspended onto a Bragg mirror or sandwiched between 2 Bragg mirrors. Strong inhibition of the Purcell factor is observed in a broad spectral range, whatever the in-plane orientation and location of the emitting dipole. We analysed these results numerically and theoretically by simulating the experimentally observed lifetime of a collection of randomly distributed emitters, showing that their average emission rate is decreased by more than one decade, both for coupled or isolated emitters.

## 1. INTRODUCTION

Since the late 80's, Photonic Crystal (PhC) were used to optimise interactions between light and matter. In most of cases, using the Bloch modes supported by the PhC, the main goal was to control the absorption [1, 2, 3], or the emission [4, 5, 6] of devices such as photodetectors, solar cells, microlasers or single photon sources. For the latter, the goal is not only to increase the emission rate in one single mode using Purcell effect, but also to avoid all parasitic emissions in order to funnel spontaneous emission (SE) in a single mode of interest [7]. In other words, the goal is to inhibit SE in all modes except one. Inhibiting SE in PhC have been studied since the seminal work of E. Yablonovitch [8], where  a 3D photonic band-gap structure is used to stifle the SE rate in semiconductors. This method allows for controlling the local density of optical modes (LDOS) at the position of the emitters and on a large spectral range. Inhibition on such sample was experimentally demonstrated [9], but, due to difficulty in elaboration of 3D structures, the measured inhibition factors, which is defined as the ratio of the SE rate in bulk material over the SE rate in the PhC structure, remained relatively low (<10). However the control of the SE rate could be reached whatever the location and the polarization of the emitter. Other approaches were also studied, where inhibition is stronger but with constraints on the emitter position or orientation. For example, using micro pillars grown on a mirror leads to an inhibition factor of 16 for dipole emitters orthogonal to the nano wire [10] but the effect was dependent on the position of the emitters (typically a quantum dot) in the plane where they have been grown. Cavities in 2D PhC was also theoretically studied, such as H1 cavities [11] exhibiting huge inhibition (>1000) but for specific wavelength and location of the dipole. Tamm plasmon structures were also used leading to an inhibition Purcell factor of 40 (the highest published experimental value to our knowledge) [12] but for a single emitter accurately located at a field intensity node. The above described approaches, as their effect on inhibition is strongly spatially dependent, are therefore (i) not well adapted for a collection of emitters and (ii) hazardous with a single emitter whose location is not easily controlled.

A trade-off between emitters' position tolerance and effective inhibition factor can be found using structures with periodicity in 1 or 2 directions. Bragg mirror cavities [13] or 2D-PhC [14, 15] were used and Purcell inhibition factors of about 3 and 15 respectively have been measured. In this paper we propose to combine a vertical and in-plane strong index patterning, the so called 2.5 D PhC approach [16, 17], to achieve SE inhibition on a whole plane for electrical dipole emitters with dipole momentum lying in this plane. More specifically, a 2D PhC will be used to remove guided mode in the membrane via photonic band gap effect as proposed in [18], and a vertical patterning will be designed to inhibit the coupling of the emitters with radiative modes.

## 2. STUDIED STRUCTURE

To perform our study, we used 3D FDTD simulations [19] to optimise the samples presented on Figure 1.

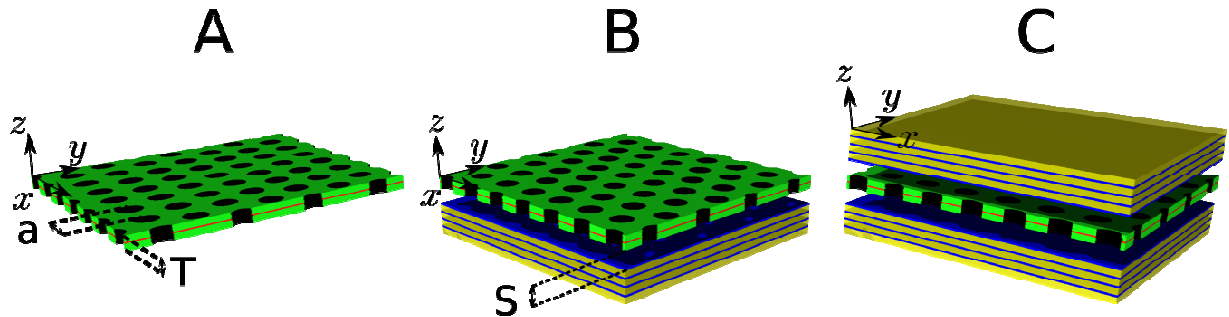

**Figure 1 : Schemes of the studied structure, A- triangular 2D-Phc membrane suspended in air, B- triangular 2D-Phc suspended on the top of a Bragg mirror, C- triangular 2D-PhC sandwiched between two Bragg mirrors.**

The first sample consists in a stand-alone 2D-PhC membrane with triangular symmetry ($C_6$). The PhC is square with a side length of 17 periods. The second sample is obtained by suspending the first one above a Bragg mirror. In the last sample the PhC membrane is sandwiched between two Bragg mirrors. The lattice parameter of the crystal is called $a$, the thickness of the membrane $T$, the air space between the membrane and the mirrors $S$. The air filling factor is set to 0.4 (in area), known to exhibit a large band gap. The mid plane of the membrane is shown in red on the figure, and corresponds to the location of the emitters. We choose a refractive index of 1 for the air and 3.5 for the material which corresponds e.g. to the index of GaAs at low temperature (~10K). All these structures feature a high index contrast patterning in both vertical and planar directions.

This property, as it allows a tight control of the 3D optical density of modes, is of prime importance for the results presented in this paper. In order to obtain a maximum inhibition factor a suspended PhC membrane is studied in this paper. However similar results can be obtained if the membrane is deposited on a low index material such as silica. In the following, seeking at maximizing the inhibition factor, the geometry of the stand-alone PhC membrane (sample A) is first optimized in order to minimize the coupling of the emitters with waveguide and free-space modes. Then, adding the bottom Bragg mirror (3.5 periods of quarter wavelength *bi-layer with optical indices n = 3.5 (e.g. silicon) and n = 1.5 (e.g. silica) [20]),* the effect of the spacer thickness S, in sample B, is investigated. For both samples, the spectral and spatial dependence of the inhibition factor will be analysed. Results on the full 2.5D structure with 2 Bragg mirrors (sample C) will be presented. Finally the effect of such structures on the emission of a collection of emitters randomly distributed in the mid-plan of the membrane will be studied.

### 3. OPTIMIZATION OF THE STRUCTURES

**2D structure: the stand alone PhC membrane**

First of all, using FDTD with periodical condition in X and Y and perfect match layer (PML) in Z (see Fig. 1 for the axis definition), we calculate the band diagram of the triangular PhC membrane for TE polarization (electric field lying in plane). In order to extract the frequency parameters we used Harminv software [21]. This software is based on harmonic inversion where the temporal signal is projected to the family of exponential decay or growth functions. This allows us to get not only the frequency of the modes, but also their quality factor, if they are above the light cone. This method is valid as soon as the mode is resonant enough (not too lossy). More quantitatively, in our case a quality factor larger than 20 was required. The membrane thickness T is chosen sufficiently small for a monomodal behaviour of the planar waveguide, in the considered range of wavelengths. The band diagram is given in Figure 2.

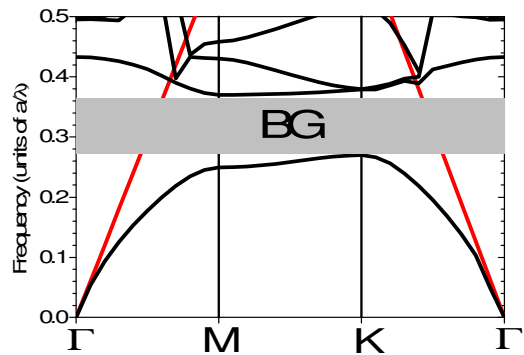

**Figure 2 : band diagram of the triangular 2D-PhC membrane for a thickness T= 0.31 *a* and a 40% filling factor**

A broad band gap is obtained, from *a/λ= 0.23* to *0.28*, far enough to achieve an efficient inhibition, on a large spectral range, for guided modes. However, one should notice that with this method, every mode with Q bellow 20 is ignored. These low Q modes include Fabry Perot membrane modes which cannot be depicted on the figure. Though, these radiative modes contribute to the LDOS of the structure and then to the emission rate of the emitters, even if they are located in the bandgap [14]. Coupling of the emitters to these free space radiative modes can be reduced by optimizing the thickness of the membrane.

For emitters lying in the mid plane of the membrane, a good way to avoid coupling with modes propagating in the vertical direction is to set the membrane thickness to $\lambda/2n$. In this case, vertical radiative modes will exhibit a node on the plane of the emitters and the coupling will be minimized. Because of the difficulty to know precisely the effective refractive index of the PhC membrane, reflectivity (transmittance) simulations were performed. The minimum of reflectivity (maximum of transmittance) give a frequency corresponding to a half wavelength thickness. This is illustrated on Figure 3 where the frequency corresponding to a "$\lambda/2n$" thickness is plotted versus $T$, along with the valence and conduction band edges.

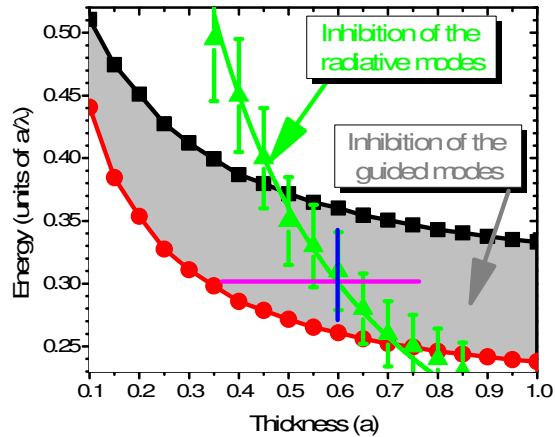

**Figure 3 : Valence band edge (red round), conduction band edge (Black Square) of the 2D-PhC and frequency corresponding to a λ/2n thickness (green triangle), as a function of the membrane thickness**

On this figure, the band gap is crossed by the "$\lambda/2n$' line, giving a range of parameters where inhibition can be obtained both for guided and radiative modes.

To validate this approach, we calculated the SE inhibition factor γ which is the inverse value of the commonly used β enhancement factor. The SE inhibition factor (γ) is calculated as the ratio of : -the steady state power exiting of the close surface of the simulation windows containing the studied structure excited with a current source -by the steady state power exiting a simulation containing the same current source standing alone in the same window but fill with a bulk $n=3.5$. FDTD simulations are performed with absorbing boundary conditions (PML) surrounding the samples. A monochromatic TE current source smaller than the spatial sampling ($a/16$) is put in the mid plane of the membrane out of any high symmetry point (see Figure 7 for precisions). As a starting point, the orientation of the dipole is chosen along Y: the effect of the polarization will be investigated later.The total emitted power is calculated for an emitting dipole located into the photonic structure and for the same dipole in the bulk material. Then γ is evaluated as:

$$\gamma_{structureX} \triangleq \frac{1}{\beta_{structureX}} = \frac{\left[\oiint \vec{S}dS\right]_{structureX}}{\left[\oiint \vec{S}dS\right]_{Bulk}} \qquad (1)$$

Where $\vec{S}$ is the Pointing vector and where the integral is evaluated on any close surface far from the structure. Calculation was done for different sets of parameters for the PhC. corresponding to the green, blue and purple lines in figure 3 and the results are depicted in figure 4.

The PhC structure is a 17 periods wide square. This size is large enough to result in side lateral losses, by optical tunnelling through the finite PhC, negligible compare to radiative losses. This can be seen on the left curve of the figure 4, where the inhibition remains at the same approximate level inside the band gap (from $a/\lambda$=0.27 to 0.33).

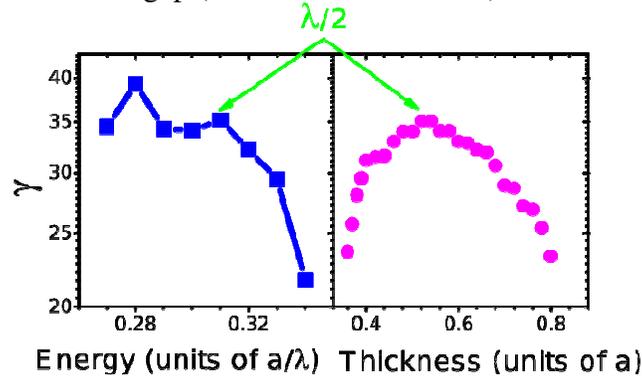

**Figure 4 : SE enhancement factor as a function of energy (blue square, left) for thickness of 0.6 a, and as a function of thickness (purple round, right) for energy of 0.3 a/$\lambda$, corresponding to the lines with same colours in Figure 3.**

To ensure better inhibition and because of imprecision of the methodology given the $\lambda/2n$ frequency we decided to calculate inhibition around our working point versus frequency of the emitters (**Figure 4** left) and the thickness of the membrane (**Figure 4** right), corresponding to the colour line on Figure 3.

It can be seen that the highest inhibition factor is obtained for parameter slightly different from the "$\lambda/2n$" ones (actually, lower energy on blue curve and smaller thickness on purple curve). Indeed, the SE rate is determined by the local density of state (LDOS) at the mid-plane of the membrane, which is decreased through the choice of a "$\lambda/2n$" thickness, but it has to be noticed that the global DOS (without taking into account the distribution of the field intensity) increases with the thickness membrane, as described in [22]. Therefore, for a given wavelength, a trade-off is obtained for a thickness slightly smaller than "$\lambda/2n$".

## 2.5 D structures

With this configuration of high index 2D-PhC "$\lambda/2n$ *thick*" membrane, we already obtain an inhibition factor as high as 35. This value results from a decrease of the LDOS at the location of the emitters:

- Due to the PBG effect, the density of guided modes vanishes.

- The optimization of the membrane thickness allows for the reduction of the LDOS for free-space modes at the vicinity of the emitters (mid-plane).

This last effect, which is due to the partial destructive interferences between the wave reflected at the membrane interfaces, can be strengthened by engineering the vertical index patterning. . A way is to suspend the membrane on top of a Bragg mirror consisting in 3.5 periods of quarter wavelength (for example at a wavelength of *3.33 a) bi-layer with optical indices n = 3.5 (e.g. silicon) and n = 1.5 (e.g. silica [20])*. As depicted in black on figure 5, the inhibition factor is calculated as a function of the spacer thickness, S, between the Bragg mirror and the PhC membrane. For small S (<λ/4), the near field coupling between the source and the Bragg guided modes leads to a decrease of γ and to a degradation of inhibition. For larger S, an oscillating behaviour is observed which is due to the interference process experimented by radiative waves between the mirror and the membrane [17]. An optimum is reached for S/λ = 0.37 (note that the 2 minima of SE rate are approximately separated by a half wavelength).

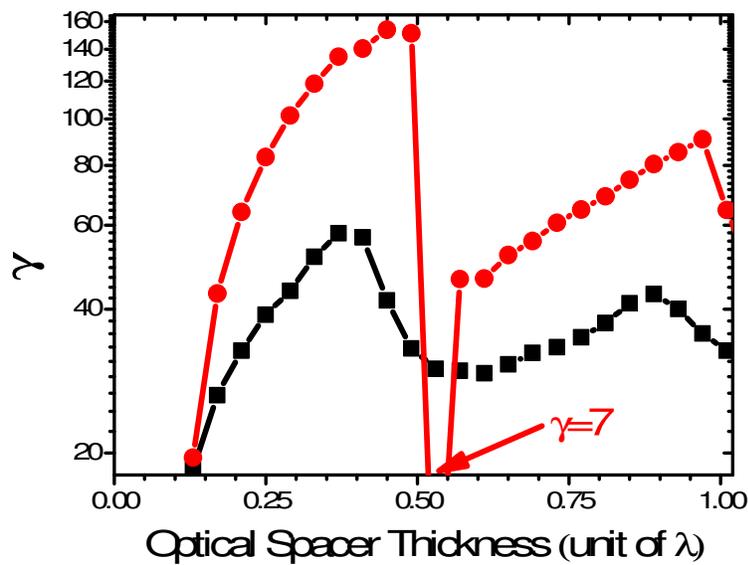

**Figure 5 : Inhibition value versus the thickness of the spacer for the optimized 2D-PhC membrane suspended on a Bragg mirror (B structure, black curve) and sandwiched between 2 Bragg mirrors (C structure, red curve)**

In addition, for large spacer (~λ) it can be noticed that γ already tends to the one without mirror meaning that the value corresponding to "infinite" spacer is reached rapidly. This behaviour has been previously observed by *Amos and Barnes* [23], and *Worthing et al.* [24] where they study the lifetime of an emitter located above a metallic mirror and separated by a dielectric spacer.

In case of sample C where the membrane is sandwiched in between 2 Bragg mirrors, the behaviour is similar and the enhancement slightly better, except for *S*=0.53 where the inhibition strongly decreases. Actually a vertical cavity is built by the 2 Bragg mirrors, leading to additional Fabry-Pérot modes that enhance the LDOS, and thus increase the Purcell factor. For this reason but also to avoid too complicated technologies; we will focus now on sample B.

For the spacer thickness, S=*0.37 λ*, corresponding to the best inhibition on this sample, the dependence of the inhibition factor on the wavelength of the emitter, for 2 orthogonal in plane polarization, is calculated (Fig. 6).

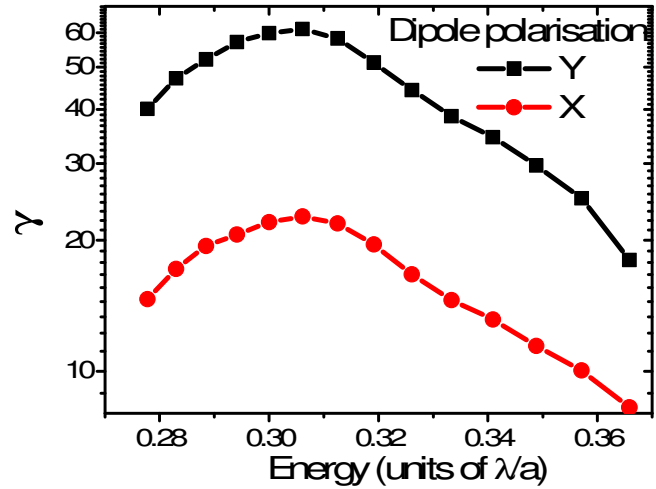

**Figure 6 : Inhibition value versus the wavelength of the emitter in the optimized 2D-PhC suspended on a Bragg structure.**

One can see that a strong inhibition can be obtained for both dipole directions (X and Y) on a broad spectrum. In addition, it is worthwhile to note that the values of the SE rate outside the band gap, for wavelength below 2.7 *a* and above 3.6 *a* (i.e. frequency outside the range 0.28-0.37 *a/λ*), correspond to a coupling with high Q band edge modes (c.f. Figure 2 and Figure 3) and cannot be calculated in reasonable time. That means that on the border of this graphics, there is a step from a Purcell inhibition regime to a Purcell enhancement regime. One can also notice that, when the dipole source vector is directed along *y*, the inhibition is lower than for *x-oriented dipole*. This will be explained with the spatial study described in the following.

In order to show if the inhibition really occurs in the whole mid-plane of the PhC, whatever the dipole in-plane location, we calculate the inhibition factor (γ) for dipole of both polarisation (*x* and *y*) as a function of the position of the dipole source. The results are depicted on Figure 7.

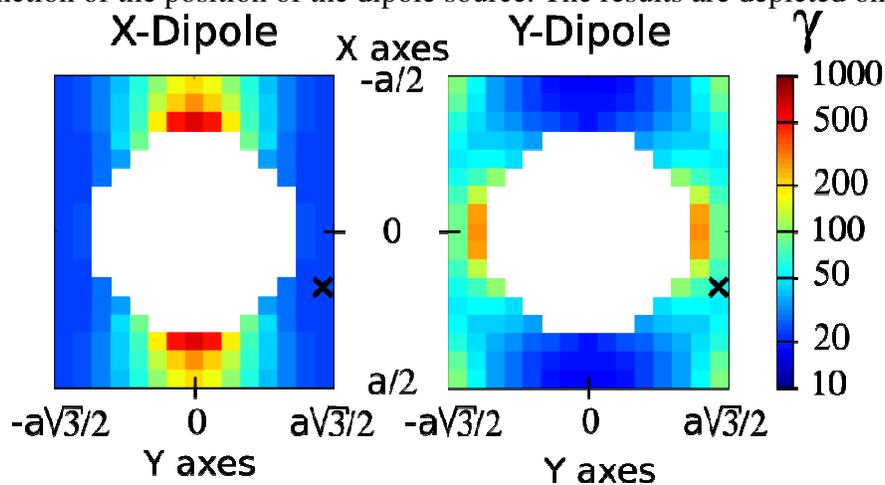

**Figure 7 : Inhibition value as a function of the position of the emitter inside the 2D-PhC cell in the optimized 2D-PhC suspended on a Bragg structure. The black crosses denote the position of the dipole in previous figures. The shapes for the other structures are similar.**

The inhibition factor remain high on the whole surface of the cell (the minimum for x-polarised dipole is 16 and for y-polarised dipole 17), showing that we succeeded to design a structure with high inhibition factor on a broad spectrum for a whole plane of emitters. One can also observe that, for each polarisation, there is one specific location where a huge inhibition is obtained ($\gamma \sim 600$ for the x-polarised dipole and ~300 for the y-polarised dipole). This is due to the continuity condition of the normal component of displacement field, called D in Maxwell equations (see for example [25]):

$$E_{n-air} = \varepsilon E_{n-material} \qquad (2)$$

In other words, the electrical energy density seen by a dipole normal to the surface can be roughly estimated to be "$\varepsilon^2$ times" weaker than for a dipole parallel to the surface as it has been seen for example in [26].

## 4. EFFECT ON THE APPARENT TIME DECAY OF A COLLECTION OF EMMITERS

In this section we analysed the role of Purcell inhibition on the lifetime of emitters embedded in the studied 2.5D structures. Keeping the parameters of sample B, the sample C is simply designed by putting an additional Bragg mirror above with the same spacer thickness (S=0.37λ) in order to prevent the presence of Fabry Perot modes. First of all, we calculated the inhibition factor, for sample A and C, as a function of the dipole location and polarization. The obtained spatial distributions are similar than the ones shown on Figure 7. However the mean values of inhibition are different. For an easy comparison, the Table 1 presents the maximum, "mean" (this is in fact 1 over the mean of β) and minimum values of the inhibition factor for the 3 structures.

|  | $\gamma_{min}$ | $\gamma_{"mean"}$ | $\gamma_{max}$ |
|---|---|---|---|
| Sample A | 9 | 21 | 417 |
| Sample B | 16 | 36 | 625 |
| Sample C | 21 | 41 | 653 |

**Table 1 : Minimum, mean and maximum values of the inhibition factor for sample A B and C for both polarizations.**

One can see a great difference for sample A on the one hand and sample C on the other hand, but only tiny difference between sample B and C. This conclusion is reinforced by figure 8 which presents, for each structure, a histogram of the occurrence of γ *in log scale (dB) in term of number of pixels of the maps shown on figure 7.*

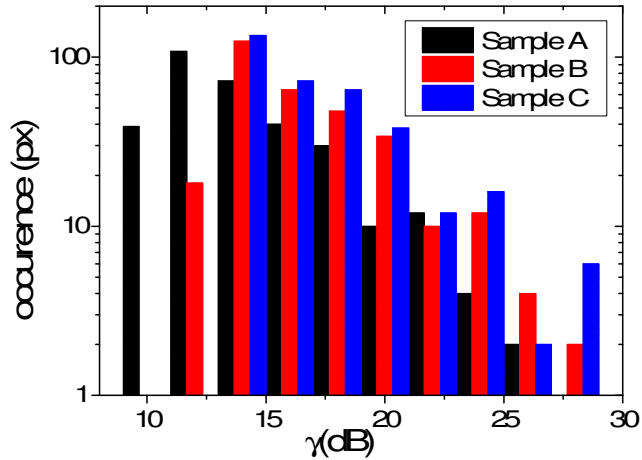

**Figure 8 : Histogram of the occurrence of the calculated 10log($\gamma$) on maps for the 3 structures, without Bragg mirror in black suspended on one Bragg mirror in red and sandwiched between 2 mirrors in blue.**

Actually, as shown on Figure 5, evanescent coupling to the mirror and coupling with Fabry-Perot modes are limiting constraints, and adding a mirror results in a modest enhancement of the inhibition.

To go further in the comparison, we evaluate the radiative lifetime of a collection of identical (in terms of life time and frequency) emitters embedded in our samples, in two cases: Isolated Emitters (IE) and linked Emitters (LE). In both cases the spectral bandwidth of the emitters is smaller than the frequency range where inhibition occurs, the wavelength is set at *3a* and the lifetime of the emitters in the bulk material is $\tau$. In the first case, emitters are supposed to be "excitonically" isolated or disconnected (isolated emitters IE). In other words, their electronic states are not coupled. This is the case for a plane of diluted quantum dots. That means that, if all emitters are excited at t=0, the power emitted by the sample will be the sum of the emitted power by each emitter that can be written as:

$$I_{IE}(t) \triangleq \frac{\iint\limits_{surface} \beta \, e^{-t\frac{\beta}{\tau}} ds}{\iint\limits_{surface} ds} \tag{3}$$

The denominator term is a normalisation factor.

The second case depicts, for example, the case of a quantum well (QW) where the excitonic states are delocalized in the mid plane of the membranes. Then at each infinitesimal time, the probability to emit is proportional to the mean value of $\beta=1/\gamma$ *on the path* of the exciton. This path is assumed (hypothesis) to cover at least a unit cell in a homogeneous way. In this case the decay of the PL intensity (the light that would be measured by a detector versus time) for these linked emitters (LE) is simply:

$$I_{LE}(t) \triangleq e^{-t\frac{\iint\limits_{surface} \beta ds / \iint\limits_{surface} ds}{\tau}} = e^{-t\frac{\beta_{mean}}{\tau}} \tag{4}$$

With these formulas we obtain the decay curves depicted on Figure 9.

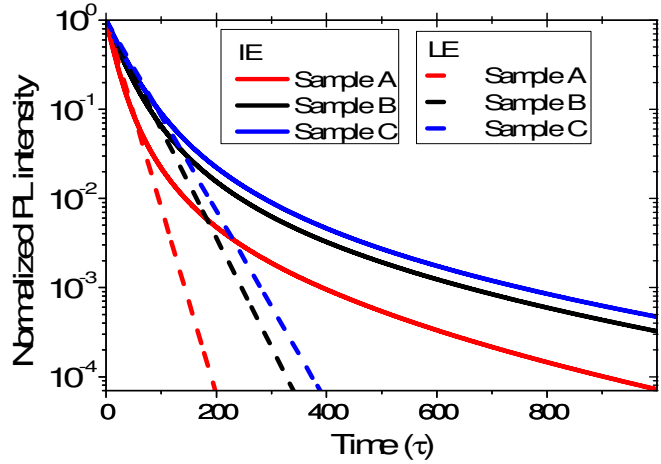

**Figure 9 : Simulated decay rate with formula (2) straight line and (3) dashed line, for sample A in black, B in red and C in blue.**

First one can observe that, in both PE and IE cases, there is a strong difference between sample A on one hand and sample B and C on the other, but tinny difference between sample B and sample C reinforcing our previous statement.

In addition, one can easily see that the decay for IE curve is not exponential. In fact, because the decay time of each pixel is different and because they are not linked like in PE case, the emission probability will change with time. In that extent it is not straightforward to define a global decay time for a whole sample. At least by taking the value of time decays at short time (for $\delta t = 10^{-2}\tau$ similar values are given using formulae from appendix), as given in Table 2, one can more easily compare the samples.

| | $\tau_{IE(0)} \dfrac{-I_{IE}(0)}{\left(\dfrac{dI_{IE}(0)}{dt}\right)}$ | $\tau_{LE(0)} \dfrac{-I_{LE}(0)}{\left(\dfrac{dI_{LE}(0)}{dt}\right)}$ |
|---|---|---|
| Sample A | 16 | 21 |
| Sample B | 28 | 36 |
| Sample C | 33 | 41 |

**Table 2 : Decay rate at short time for sample A B and C in case of isolated and linked emitters. (To see theoretical formulas for these values please see the appendix).**

Considering inhomogeneity in the PE case, at short time the pixels with the highest $\beta$ are more likely to emit first. This will lead to a smaller apparent decay time for short time compare to the IE case and thus a lower apparent inhibition factor.

The values presented in table 2 correspond to numerical application of the theoretical formulas calculated in appendix. These formulas generalise the result and can be summarized as:

$$\frac{\tau_{IE(0)}}{\tau_{PE(0)}} \leq 1 \quad and \quad \tau_{IE(t > \tau\gamma_{\min})} \approx \tau\gamma_{\min} \geq \tau_{PE(t > \tau\gamma_{\min})} \tag{5}$$

This means that, *for the short time, the highest apparent inhibition is for PE, while for long time, it is for IE.* This can be interpreted as, if for PE the decay time is always the same during the decay, for IE it changes from preponderant **low inhibition** at short time (the fast emitters emit the first) to preponderant **high inhibition** at long time (when fast emitter emitted, only slow emitters remain).

To summarize all this part, sample A will exhibit an apparent short time inhibition factor of 16 for IE and 21 for PE, adding a mirror the inhibition reach respectively 28 and 36. The second mirror increases a little more the inhibition to reach respectively 33 and 41. For sample B and C, the mean apparent inhibition values are comparable to the one experimentally induced for a Tamm Plasmon structure [12], but can occurs not only for an emitter located at a specific position but also for a spatially distributed collection of emitters.

## 5. CONCLUSION

To conclude, with FDTD simulations, using a 2.5D PhC approach, we succeed in designing a structure exhibiting inhibition for any position of a dipole on the mid plane of a PhC membrane. This inhibition reaches a factor upper than 33 for isolated emitters and 41 for linked emitters on a broad spectrum, for any direction of dipole in the plane of the PhC. This value is competitive with the highest experimental inhibition factor from the literature (40 [12]) but can be applied on a whole plane of emitter and is not limited to a single emitter. Then we analysed how the inhomogeneity of inhibition inside the membrane mid-plane may affect experimental life time measurement in case of connected emitters and disconnected emitters. This lead to confirm that inhibition will be measurable on this kind of sample on any kind of emitters.

To go further and achieve even stronger inhibition 2 paths should be followed. First of all, optimisation of the filling factor can allow to slightly increasing the inhibition factor. In addition, studying the evanescent coupling to back (and top) mirror and comparing this coupling in case of Bragg mirror, metallic mirror or even photonic crystal mirror could be a way to improve the proposed structure.

In terms of applications, it has to be noticed that the inhibition remains very high close to the band edge of the photonic crystal band gap. Considering that it is possible to change locally this band edge frequency by changing one parameter of the crystal (hole radius for instance); it would be possible to make a photonic heterostructure, where the crystal will inhibit emission everywhere but at a specific location avoiding emission toward unwanted modes. It can be used to make such devices as very low threshold laser or single photon source with high efficiency.

## 6. ACKNOWLEDGMENT

The authors are grateful to Guillaume Gomard for the help to improve the manuscript. This work is supported by the French National Research Agency (ANR) through the Nanotechnology program (CAFE project). Romain Peretti also thanks Pierre Clare for fruitful discussions.

# 8. APPENDICES

## Bragg mirror reflectivity at normal incidence

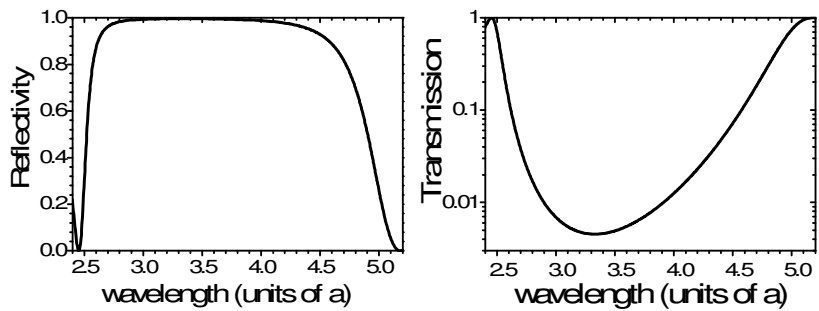

**Figure 10 : reflectivity spectrum of the Bragg mirror at normal incidence**

## Theoretical derivation of measured decay time

The values presented in table 2 correspond to the theoretical value given for isolated emitters by:

$$\tau_{IE(t)} = \frac{-I_{IE}(t)}{\left(\dfrac{dI_{IE}(t)}{dt}\right)} = \tau \frac{\iint\limits_{surface} \beta\, e^{-t\frac{\beta}{\tau}}\, ds}{\iint\limits_{surface} \beta^2\, e^{-t\frac{\beta}{\tau}}\, ds} \qquad (6)$$

And for linked emitters by:

$$\tau_{LE(t)} = \frac{-I_{LE}(t)}{\left(\dfrac{dI_{LE}(t)}{dt}\right)} = \tau \frac{\iint\limits_{surface} ds}{\iint\limits_{surface} \beta\, ds} = \tau \beta_{mean} \qquad (7)$$

Comparing both we get:

$$\frac{\tau_{IE(t)}}{\tau_{LE(t)}} = \frac{\iint\limits_{surface} \beta\, e^{-t\frac{\beta}{\tau}}\, ds}{\iint\limits_{surface} \beta^2\, e^{-t\frac{\beta}{\tau}}\, ds} \frac{\iint\limits_{surface} \beta\, ds}{\iint\limits_{surface} ds} \qquad (8)$$

That can be simplified at t=0 in :

$$\frac{\tau_{IE(0)}}{\tau_{LE(0)}} = \frac{\left(\iint\limits_{surface} \beta\, ds\right)^2}{\iint\limits_{surface} \beta^2\, ds \iint\limits_{surface} ds} \le 1 \qquad (9)$$

Using Cauchy Schwarz inequality with function "$I$" and $\beta$.
In addition, for long time

$$\tau_{IE\left(t > \frac{\tau}{\beta_{min}}\right)} \approx \frac{\tau}{\beta_{min}} \qquad (10)$$

Because all the Purcell factor are positive and smaller than 1 in case of inhibition. This means that at longer time, the exponential decay time equivalent value for linked emitter will be smaller than for isolated emitter as it can be seen on Figure 9.